# Data-efficient 4D-STEM in SEM: Beyond 2D Materials to Metallic Materials

Ujjval Bansal[1], Amit Sharma[2], Barbara Putz[2,3], Christoph Kirchlechner[1], Subin Lee[1,*]

[1]Institute for Applied Materials, Karlsruhe Institute of Technology, 76131 Karlsruhe, Germany

[2]Empa, Swiss Federal Laboratories for Materials Science and Technology, 3603 Thun, Switzerland

[3]Department of Materials Science, Montanuniversität Leoben, 8700 Leoben, Austria

*corresponding author: subin.lee@kit.edu

**Abstract**

Four-dimensional scanning transmission electron microscopy (4D-STEM) is a powerful tool that allows for the simultaneous acquisition of spatial and diffraction information, driven by recent advancements in direct electron detector technology. Although 4D-STEM has been predominantly developed for and used in conventional TEM and STEM, efforts are being made to implement the technique in scanning electron microscopy (SEM). In this paper, we push the boundaries of 4D-STEM in SEM and extend its capabilities in three key aspects: (1) faster acquisition rate with reduced data size, (2) higher angular resolution, and (3) application to various materials including conventional alloys and focused ion beam (FIB) lamella. Specifically, operating the MiniPIX Timepix3 detector in the event-driven mode significantly improves the acquisition rate by a factor of a few tenths compared to conventional frame-based mode, thereby opening up possibilities for integrating 4D-STEM into various *in situ* SEM testing. Furthermore, with a novel stage-detector geometry, a camera length of 160 mm is achieved which improves the angular resolution amplifying its utility, for example, magnetic or electric field imaging. Lastly, we successfully imaged a nanostructured platinum-copper thin film with a grain size of 16 nm and a thickness of 20 nm, and identified annealing twins in FIB-prepared polycrystalline copper using virtual darkfield imaging and orientation mapping. This work demonstrates the potential of synergetic combination of 4D-STEM with *in situ* experiments, and broadening its applications across a wide range of materials.

**Graphical abstract**

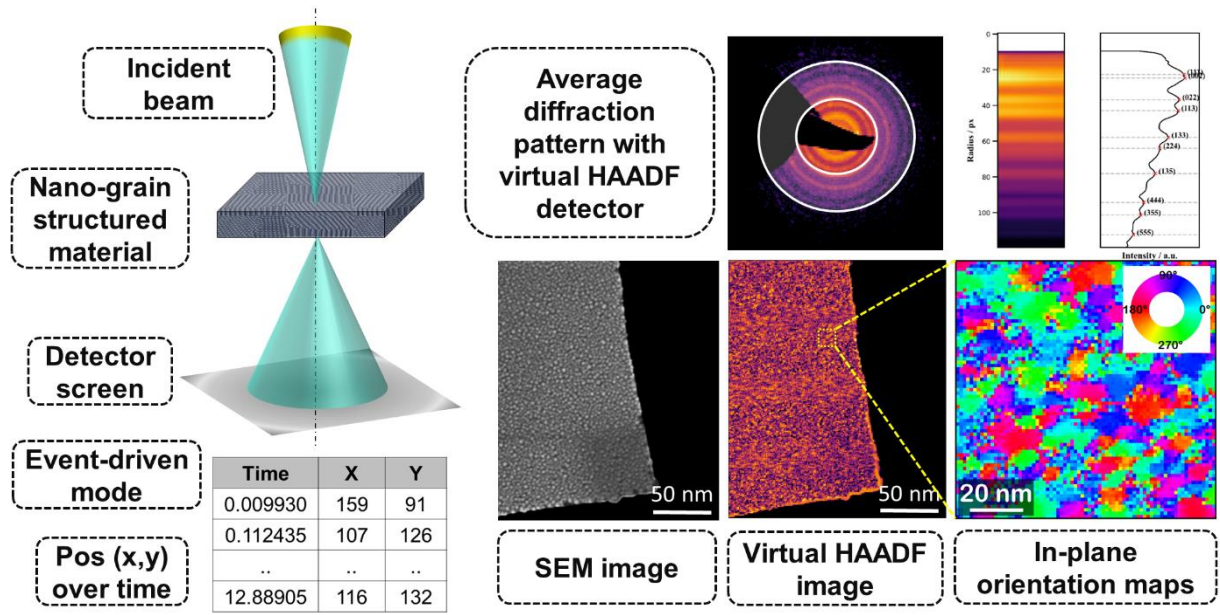

# 1. Introduction

Four-dimensional scanning transmission electron microscopy (4D-STEM) is a technique involving the collection of 2D images in both real and reciprocal space, respectively, through the scanning of a converged electron beam over an electron-transparent sample. This method yields a 4D dataset, hence its nomenclature, 4D-STEM. It is essential to clarify that this dimension does not pertain to time; rather the resulting 4D dataset contains both spatial and diffraction information. The applications of 4D-STEM measurements in material science are extensive, particularly in characterizing crystal orientation mapping, dislocations, strain mapping, or field information such as electric or magnetic fields, with high spatial and angular resolutions [1,2].

Although 4D-STEM has been predominantly developed for and utilized in conventional TEM, efforts have been made to implement it in scanning electron microscopy (SEM). This accessibility, combined with the ease of installation in existing microscopes, cost-efficiency compared to TEM, and operational simplicity, has stimulated interests within electron microscope communities toward adopting 4D-STEM in SEM, which can potentially expand 4D-STEM applications to a wider community. In theory, any SEM equipped with an on-axis diffraction detector, for example direct electron or pixelated detector, can perform 4D-STEM operations [3–6]. Schweizer et al. showcased an interesting approach to capturing diffraction patterns from gold and silicon on a fluorescent screen by utilizing a complementary metal oxide semiconductor (CMOS) camera [3]. This is extended further by Caplins et al. by developing a new STEM detector [7]. It offers a hybrid approach, where transmitted electrons hit the scintillator screen, converting electrons to photons, and are imaged by a digital micromirror device (DMD). For a diffraction pattern, the mirrors in DMD tilt towards the CMOS camera and for real space imaging, integrating photodetector is used. Furthermore, recently pixelated detectors dedicated to 4D-STEM operations with improved quantum efficiency have been developed [8].

The energy of the incident beam in SEM is typically limited to 30kV, which is considerably lower compared to TEM. Achieving atom-resolved imaging requires sub-Angstrom resolution [9], typically attained through higher electron beam convergence angle and energies in (S)TEM together with recent advances in spherical aberration correctors. This level of resolution surpasses the capabilities of SEM, but sub-nanometer resolution [10] can be achieved in SEM which is sufficient to resolve a unit cell [11]. Additionally, lower energy leads to an increase in scattering cross-section, which amplifies diffraction intensities. Another benefit is reduced beam damage, more specifically knock-on damage on metal, ceramics or inorganic materials. These factors make it an ideal tool for imaging beam-sensitive materials such as zeolites [12,13] carbon nanotubes [14]. Considerable research efforts have also been directed towards investigating graphene layers [7,15,16] encompassing monolayers and bilayers, with emphasis on revealing their texture, or phase maps using 4D-STEM in SEM. Denninger et al. [17] presented the application of 4D-STEM in revealing the dislocations in bilayer graphene in SEM and compared it with results obtained from TEM. The improved contrast due to stronger scattering and lower acquisition time is an advantage. Müller et al.'s recent work [18], showcases the single crystallinity of $MoS_2$ produced via gold-mediated exfoliation and mapping properties within a $C_{60}/MoS_2$ van der Walls heterostructure, underscores the utility of this technique. Furthermore, combining this technique with *in situ* experiments has also shown promise. Denninger et al. [19] illustrated the simultaneous monitoring of dewetting, grain coarsening and the development of [111] texture in gold thin films by the *in situ* heating stage in SEM in combination with BSE imaging and low-energy nano-diffraction (LEND). Another fascinating example shown was the crystallization of amorphous Si layers via metal-induced layer exchange shown using *in situ* heating and LEND setup (transmission imaging) [20]. Given the integration of *in situ* with transmission mode in SEM, it demonstrates the potential of combining it with 4D-STEM in the future.

Unlike conventional TEM which operates at voltages ranging from 100-300 kV, SEM typically operates at lower voltages, around 20-30 kV. These lower voltages enhance image contrast for low atomic number components or low-density materials as demonstrated in the examples above. However, exploring metallic materials or conventional focused ion beam (FIB)-prepared samples in STEM mode within SEM remains challenging. Additionally, the slow acquisition speed of SEM has limited its application and often results in artifacts, such as severe carbon contamination. Our goal is to acquire crystallographic information from metallic samples in transmission mode at these lower accelerating voltages while achieving faster acquisition speeds. To address the challenge of obtaining transmission electrons from metallic samples, thin specimens with sufficient electron transparency were fabricated.

This paper discusses the implementation of 4D-STEM in SEM, pushing its capabilities in three key areas: faster acquisition rate with reduced data size, higher angular resolution, and broad applicability to various materials including metallic materials. Firstly, we established protocols for imaging diffraction patterns in event-driven mode instead of frame mode and investigated the influence of SEM beam parameters on its operation. Additionally, we achieved improved angular resolution by utilizing a novel stage-detector geometry with a camera length of 160 mm. This study utilized two sets of specimens to demonstrate these advancements: 20 nm thick Pt-Cu thin films, and polycrystalline Cu containing annealing twins prepared by Ga$^+$ FIB. Our results highlight the enhanced potential of 4D-STEM in SEM for comprehensive materials analysis.

## 2. Materials and Methods

### 2.1. Sample fabrication and Microstructural characterization

Two samples were used in this study. Firstly, thin films were deposited by direct current (DC) magnetron sputtering, directly on a TEM grid with a base pressure of $7 \times 10^{-5}$ Pa and a working pressure of $5 \times 10^{-1}$ Pa with 25 sccm of Ar flow. A 5 nm Cu layer was deposited at room temperature from a 2" circular Cu target (purity 99.99%) using a DC power of 3W. The estimated Cu deposition rate was 0.006 nm/s with a deposition time of 14 min. After deposition, the film was annealed in vacuum at 650°C for 8 hours to form Cu nanoparticles via dewetting. Subsequently, 20 nm Pt thin films were deposited from a 3" circular Pt target (purity 99.99%) using a DC power of 8.5 W. The estimated Pt deposition rate was 0.04 nm/s with a deposition time of 8 min. The microstructure examination of the thin film was characterized using a (S)TEM (Titan Themis, Thermo Fisher Scientific) operating at 300 kV, equipped with a probe corrector. The high-angle annular dark-field (HAADF) STEM were carried out in STEM nanoprobe mode. The elemental maps were obtained using energy-dispersive X-ray spectroscopy (EDS) using a Super-X detector.

A second set of samples from polycrystalline Cu was prepared directly from a PELCO® FIB lift-out grid. The initial thickness of grids was ~35 μm. Using Ga$^+$ FIB milling (Crossbeam 550L, Zeiss) the sample was thinned down to less than 100 nm. Coarse milling was conducted at an acceleration voltage of 30 kV, starting with an initial current of 1.5 nA, and then 700 pA. Fine milling was performed at 30 kV and 100 pA, followed by the cleaning step at 30 kV and 20 pA.

### 2.2. Pixelated detector for 4D-STEM in SEM

A MiniPIX TPX3 detector (Advacam) with a Si sensor chip (100 μm thick), featuring a resolution of $256 \times 256$ pixels with a pixel size of 55 μm was employed to detect the transmitted electrons. The minimum detectable energy of the camera is 3 kV and the maximum readout speed is 2.35 million hits per second. Further, each pixel can independently measure events with a precision of 1.6 nanoseconds. The operational parameters for the detector, such as acquisition time, frame count, and output file, are programmable via the PIXET PRO software, and additionally via Python scripts.

We have designed the in-house assembly for 4D-STEM characterization. To ensure the temperature stability of the detector, a heat sink made from aluminium alloy was utilized, and connected to a water chiller maintaining a temperature of ~20°C. The sample was positioned above the detector screen as illustrated in Fig. 1(a). Both the sample holder and the detector were attached to a z-axis substage mounted on a base plate. The whole setup was installed in an SEM (Merlin, Zeiss) with a field emission gun (FEG) source. The in-chamber setup is shown in Fig. 1(b), and the schematic is presented in Fig. 1(c). Given that the SEM stage is positioned at zero tilt, this setup is referred to as the "zero-tilt" configuration.

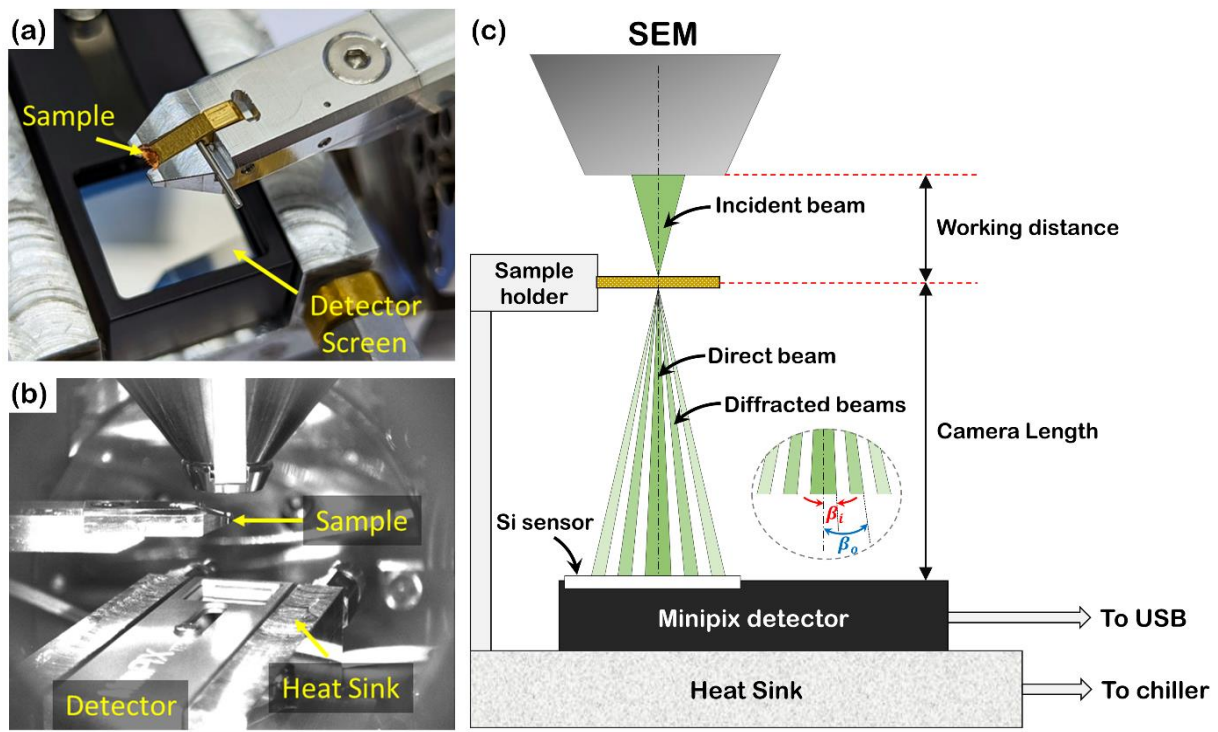

Fig. 1: (a) Picture showing a sample over the detector screen, (b) in-chamber configuration of Zeiss Merlin® during data acquisition, and (c) schematic of 4D-STEM operation at the zero-tilt configuration, inset showing inner ($\beta_i$) and outer ($\beta_o$) semi-angles of a detector.

### 2.3. Data acquisition and post-processing

The data acquisition is crucial not only for considerations of storage capacity, impacting operational efficiency but also for subsequent post-processing for developing virtual images. For current investigations, an event-driven approach for collecting the diffraction signals was adopted. Unlike conventional 4D-STEM techniques that record diffraction patterns at each real-space position, known as "frame-mode", this approach records events occurring on the screen over time. For instance, corresponding to one pixel in real space, if 100 events have occurred on the detector screen, only 100 data values are stored instead of 65,536 (=256 × 256) values corresponding to one frame. This significantly reduces the storage space requirements, which typically amount to several gigabytes for conventional 4D-STEM measurements in frame-mode.

The data in "event mode" is stored in csv format, encompassing parameters such as Matrix Index, Time of Arrival ($ToA$), Time of Threshold ($ToT$), Fast Time of Arrival ($FToA$), and Overflow. The ($x$, $y$) coordinates refer to the position where the transmitted electron hits the detector screen, and are derived from the matrix index using the formulas:

$$x = Matrix\ Index\ \%\ 256$$

$$y = Int(Matrix\ Index/256)$$

The time corresponding to each event, in nanosecond resolution, is estimated by:

$$Time\ (ns) = ToA \times 25 - FToA \times 1.5625$$

After sorting the data with time, the data is further segmented based on the line time corresponding to each line in the real-space scanning in SEM. This approach allows processing of the data to generate a diffraction pattern corresponding to each pixel in real space.

The Python package LiberTEM is utilized for the post-processing of the 4D dataset [21]. Its documentation is accessible on GitHub, offering tools for processing, analyzing, and visualizing 4D STEM data. With this library, a virtual detector was utilized to generate virtual images. For example, employing an annular ring detector enables the creation of virtual annular dark field images.

## 3. Microscope parameter optimization

### 3.1. SEM parameter optimization

This section focuses on the effects of varying SEM parameters, specifically beam current, dwell time, and camera length. By systematically exploring these parameters, our objective is to fine-tune imaging conditions and augment the efficiency of the 4D-STEM.

#### 3.1.1. Effect of beam current

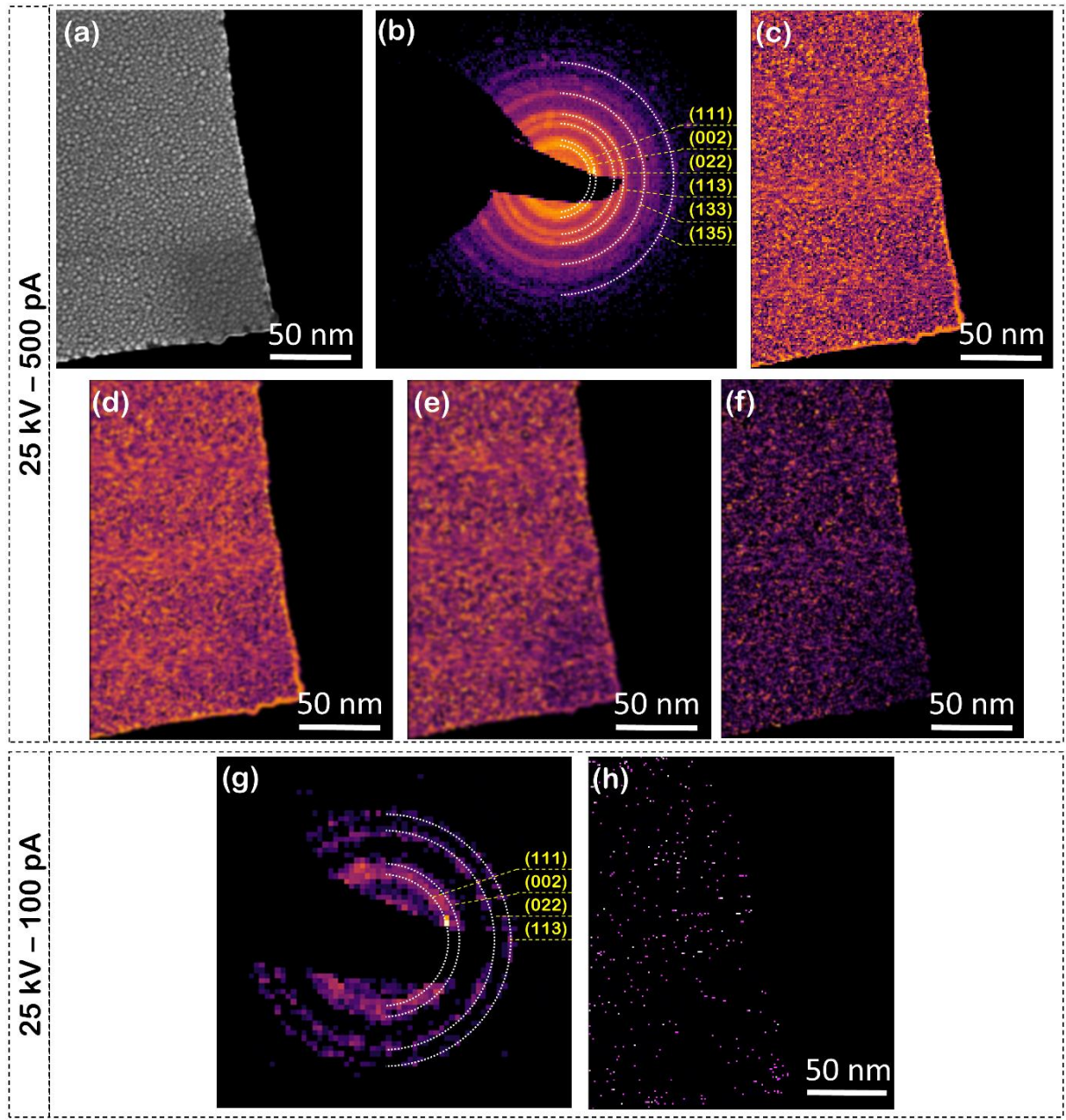

Fig. 2: (a) SEM image of a Pt-Cu thin film acquired at 25 kV and 500 pA using a conventional Everhart–Thornley (ET) secondary electron detector, (b) average diffraction patterns (indexed) and dark region corresponding to a Mo-beam stopper, (c) virtual low angle annular dark field (vLAADF) image covering 29–48 mrads, (d) Gaussian filtered LAADF image, (e) Gaussian filtered virtual medium angle annular dark field (vMAADF) image covering 50–86 mrads, (f) Gaussian filtered virtual high annular dark field (vHAADF) image covering 87–175 mrads, (g) average diffraction patterns (indexed) obtained at 100 pA, and (h) virtual ADF image using a detector covering a detection angle ranging from 30 to 83 mrad.

Fig. 2(a) displays the SEM image of Pt-Cu thin film taken at 25 kV and 500 pA using an Everhart–Thornley (ET) secondary electron detector. The reason for collecting data in such conditions is to take advantage of the fact that Pt exhibits stronger contrasts with respect to vacuum with a maximum scattering angle of 0.52 radians at 25 kV [22]. The total time to acquire the image was 13.1 seconds (for 250 × 250 pixels in real space) with the dwell time at each pixel of 209.8 µs. The diffraction patterns produced are presented in Fig. 2(b) and the visible diffraction rings were indexed according to FCC crystal structure. Please refer to the Appendix A1 for detailed instructions on the indexing procedure. To block the direct transmitted beam, Mo beam blanker were used, as shown in Fig. 2(b). This not only prevents oversaturating the detector during scanning but also improves the acquisition rate by limiting the huge amount of signals from the transmitted electron beam as the acquisition rates of a direct electron detector are mostly limited by its bandwidth. From Fig. 2(b), it is evident that higher-order Laue zones (HOLZ) are readily achieved at a camera length of 19 mm, which can be utilized for precise lattice parameter measurements as demonstrated in Si with 30 kV electron beam [23]. This is a direct benefit of using lower beam energies in SEM, which leads to a smaller radius of the Ewald sphere.

Using the LiberTEM python library, virtual images (based on the detector size and shape) were reconstructed. Fig. 2(c) displays the virtual low-angle annular dark field (vLAADF) image developed using the virtual annular detector that covers a range of scattered electrons at an angle between 29–48 mrads. The image replicates the actual SEM-acquired image. The precise alignment of interfacial features in the virtual image with the real image highlights the accuracy of the technique. Since the background noise in a diffraction data set is also reflected in the virtual image, a Gaussian filter is used to suppress the noise from the reconstruction process. Fig. 2(d) shows the vLAADF image after Gaussian filtering. Using a virtual detector spanning an angular range of 50–86 mrads, and 87–175 mrads, and overlaying a Gaussian filtering, vMAADF and vHAADF images were developed, as depicted in Fig. 2(e) and (f) respectively.

Next, we reduce the electron dose by reducing the beam current to 100 pA. The hits at detector screen was reduced from 0.1 million hits/sec at 500 pA-25kV to 1800 hits/sec at 100 pA-25kV. The diffraction patterns were obtained from the same region of Pt-Cu thin film as shown in Fig. 2(a) at reduced beam current and are displayed in Fig. 2(g). The rings corresponding to lower-order planes are faintly visible, however, higher-order planes are undetected. The beam current at 100 pA is sufficiently low to result in lower signals compared to background noise. The virtual annular dark-field (vADF) image, in Fig. 2(h) was generated using an annular virtual detector with an inner radius of ~30 mrad and an outer radius of ~83 mrad. However, the resulting vADF image lacks visible contrast due to extremely low levels of diffracted signals. This shows there are limitations to beam current reductions, considering other factors like beam alignment, and lens aberrations are as minimal, as possible. Typically, beam currents in SEM can go as high as 7 nA (hits crossing saturation limit that is 2.35 million hits/sec), however in the present investigation we work around 500 pA, aiming for better spatial resolutions and a higher acquisition rate as higher beam current causes more events leading to early saturation.

### 3.1.2. Influence of dwell time

The dwell time here refers to the amount of time that the electron beam stays at each pixel. The control of this parameter is instrument-specific, for instance in Zeiss SEMs, it can be controlled by scan speed. For a faster acquisition of data, the dwell time is a critical parameter and needs to be optimized. However, it should not compromise the retrieval of desired information to attain fast acquisition speeds.

Fig. 3 (a1-a4) shows the SEM image of the Pt-Cu thin film at higher magnification, scanned at different dwell times. It is important to emphasize that we scanned different areas because the scanning beam can contaminate the region during the acquisition. The dwell time at different scan speeds and the corresponding acquisition time (time to complete 250 × 250 pixels in real space) is provided in Table 1.

**Table 1:** Dwell time and corresponding acquisition time in SEM to record an image of 250 × 250 pixels

| Dwell time (ms) | Acquisition time to capture an image with 250 × 250 pixels (s) |
|---|---|
| 1.68 | 103 |
| 0.84 | 52.5 |
| 0.42 | 26.2 |
| 0.21 | 13.1 |

The average diffraction patterns for each sample are presented in Fig. 3 (b1-b4). Notably, even at shorter dwell time/faster acquisition, the diffraction rings remain distinctly visible. The vLAADF (29–75 mrads) and vHAADF (80–175 mrads) images are generated from each dataset to replicate the real images acquired through SEM scans. Fig. 3(c1-c4) and (d1-d4) display the Gaussian filtered vLAADF and vHAADF images.

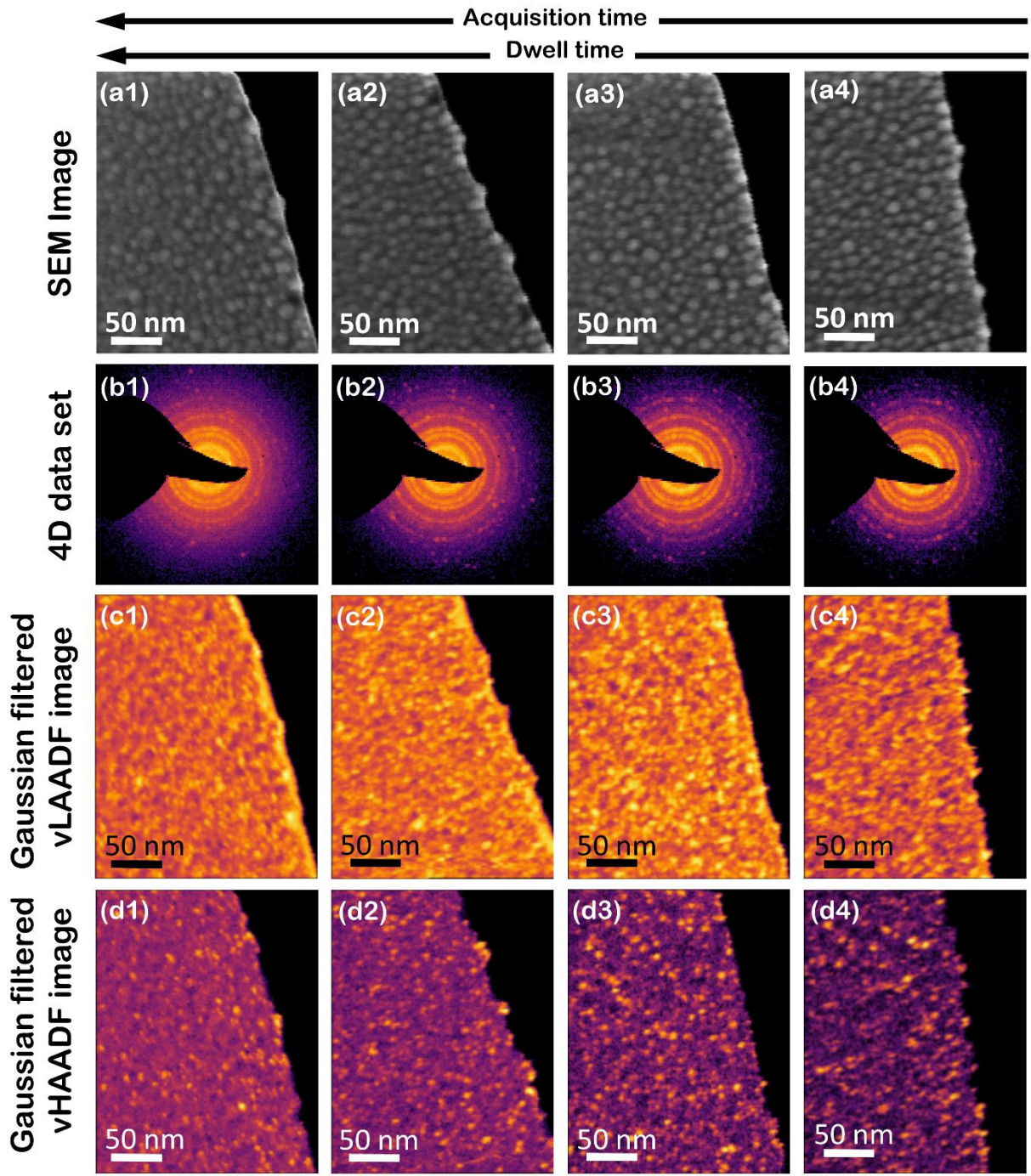

Fig. 3: SEM images using ET detector of Pt-Cu thin film acquired at 25 kV and 500 pA at a dwell time (ms) of (a1) 1.68, (a2) 0.84, (a3) 0.42, (a4) 0.21, (b1-b4) average diffraction patterns, (c1-c4) Gaussian filtered vLAADF image covering 29–75 mrads, and (d1-d4) Gaussian filtered vHAADF image covering 80–175 mrads.

To summarize the effect of dwell time, we achieved a faster acquisition time of 13.1 sec without distorting the sample interface, indicating that the integrity of the information is maintained. Further reductions in dwell time are possible, provided the desired information is preserved, for instance in the present study, as demonstrated by the in-plane orientation maps, in section 4.1, showing nano grains at a scan acquired with a dwell time of 0.21 sec.

### 3.1.3. Camera length optimization

Recent advancements in 4D-STEM enabled the precise measurement of the field properties of the materials including electric field, magnetic field and atomic potential [2]. Detecting the small angular deflections induced by the magnetic field necessitates a substantial camera length [24,25]. While conventional TEM typically adjusts camera length using an intermediate lens, our approach involved testing a high camera length setup achieved by tilting the stage to 70 degrees and positioning the detector at the bottom of the SEM chamber. Fig. 4(a) illustrates the high-tilt configuration employed in SEM to maximize the camera length. This configuration allows spreading the disk (diffraction) over a large number of pixels, particularly useful for capturing detailed diffraction spots of lower-order planes.

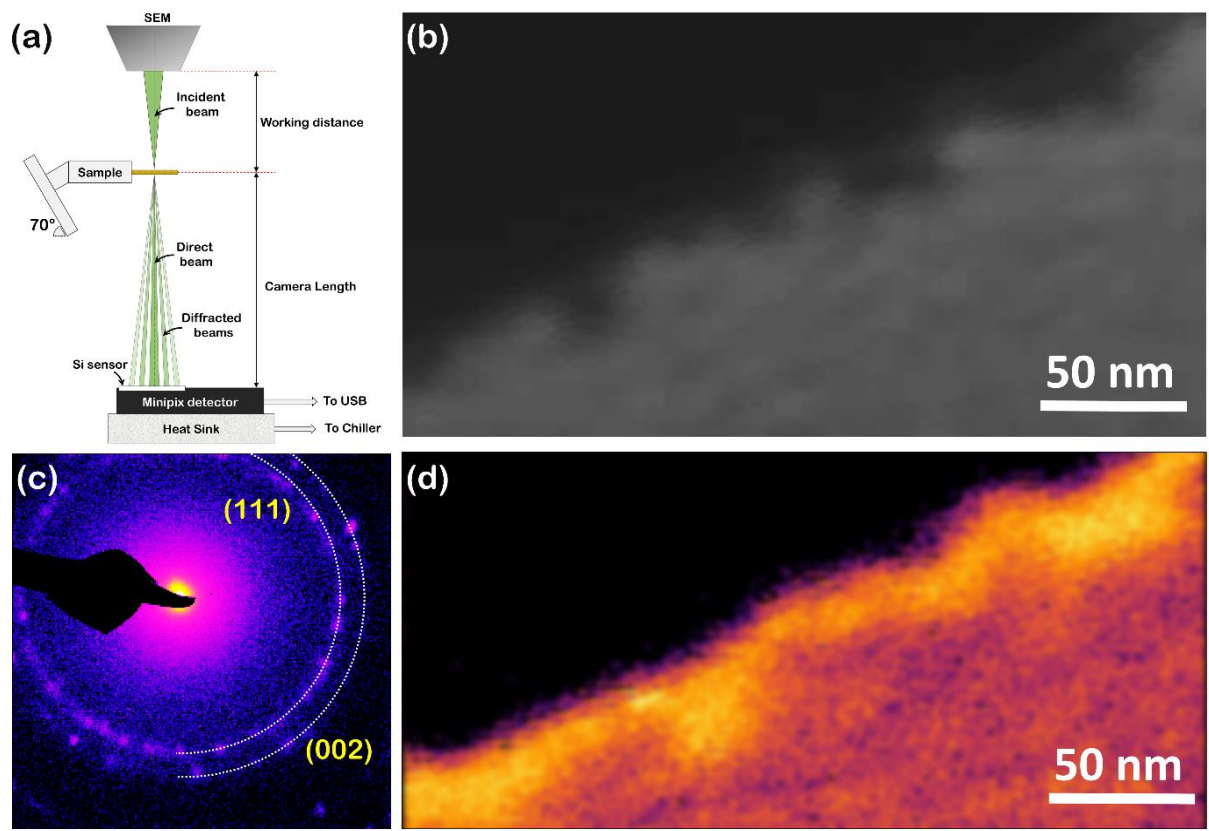

Fig. 4: (a) Schematic of 4D-STEM operation at high-tilt (70°) configuration, (b) SEM image of Pt-Cu thin film at 20 kV, 500 pA and at dwell time of 0.2 ms, (c) Average diffraction patterns (indexed), and (a) Gaussian filtered vLAADF covering the range of 32–50 mrads.

Fig. 4(c) shows diffraction patterns obtained from a Pt-Cu thin film (in Fig. 4(b)) under high-tilt conditions. Without any pixel binning, the diffraction spots corresponding to lower order planes ((111) and (002)) are well resolved, which is not visible at zero-tilt configurations (refer to Fig. 3(b)). With a high-tilt configuration, a camera length is increased by 8.5 times (~161 mm). In terms of angular resolution defined by the pixel size of the detector, one pixel in high tilt configuration is ~0.34 mrads and is ~2.89 mrads in zero-tilt configuration (19 mm of camera length). Further, the vLAADF image is developed by a virtual detector spanning over an angular range of 32–50 mrads. The effective noise reduction was achieved by employing a Gaussian filter (sigma = 3.0), enhancing image quality, while preserving the essential features. The refined image is shown in Fig. 4(d).

## 4. Case study

Using the optimized electron beam and detector parameters established in Section 3, we demonstrate the application of 4D-STEM in SEM on two material systems. First, we apply the technique to map the crystalline orientation of the previously discussed Pt-Cu thin film, and second, to FIB-prepared polycrystalline copper.

### 4.1. Measuring particle size in thin films

The HAADF-STEM image of Pt-Cu thin film in Fig. 5(a) acquired at 300 kV displays the nano-sized grains. The estimated grain size is measured to be 16.0 ± 2.5 nm and the size distribution plot is shown in Fig. 5(b). The STEM-EDS maps, in Fig. 5(c) suggest the Cu clusters in Pt grains.

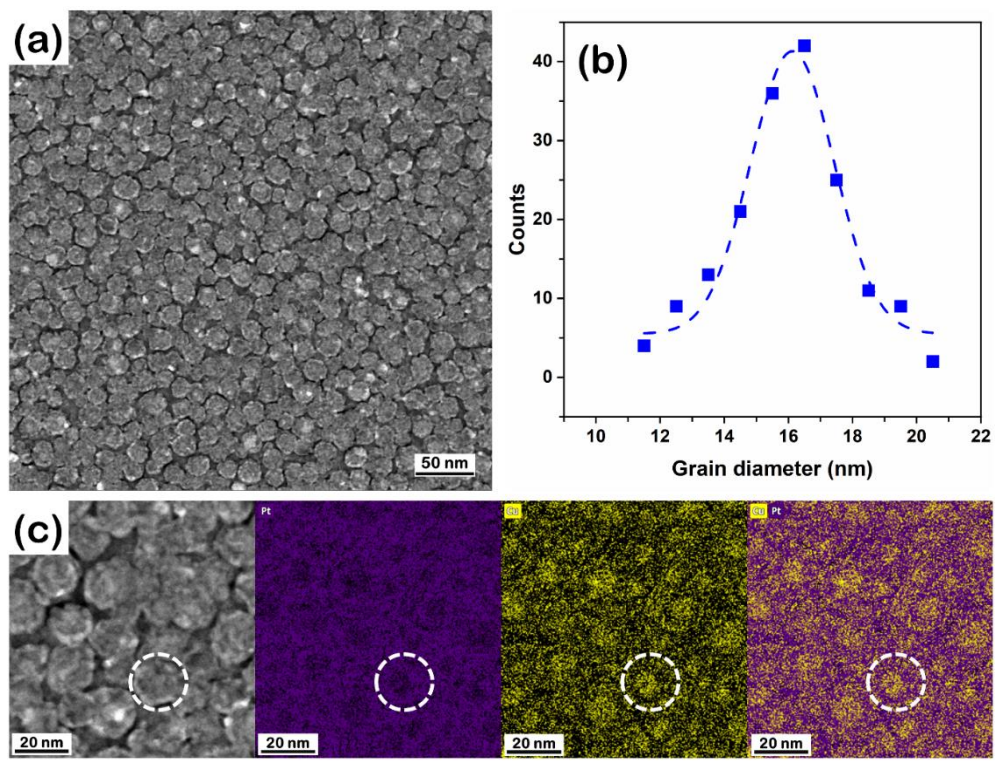

Fig. 5: (a) HAADF-STEM image of Pt-Cu thin film, (b) Histogram of grain size with a mean diameter of 16.0 ± 2.5 nm, and (c) STEM EDS map of Pt and Cu showing Cu clusters in Pt grain.

To estimate the grain size from the 4D-STEM data obtained, a Fourier analysis was applied within a ring-shaped region (marked in Fig. 6(a)) to derive an in-plane orientation map. The average diffraction patterns in Fig. 6(a) were cropped to a size of 75 × 75 pixels (real space), from the scan data acquired at a dwell time of 0.21 ms in Fig. 3. It is important to emphasize here that each pixel of the real space scan stores information corresponding to crystal structure and its orientation, that can be extracted, for instance in this case, using a radial Fourier analysis, which is challenging in traditional SEM. In this analysis, the phase angle was considered as the angle of the diffracted spot relative to the symmetric axis of the average diffraction patterns. Utilizing the Fourier coefficients in the Fourier-spaced data, absolute (using amplitude) and phase (using phase) maps can be generated. One of the advantages of constructing absolute maps is to determine the crystallinity order within the amorphous regions [26]. Importantly, for crystalline materials, the critical aspect lies in the information extracted from the phase of Fourier coefficients. The in-plane orientation map can be constructed from the first order of the Fourier coefficient, as displayed in Fig. 6(b). This highlights the ability of the technique to gather information at faster (or shorter) dwell time and low magnifications and extract intricate details, which can in future be applied for various analyses, for example, dislocation imaging or strain mapping. Further, the estimated mean grain size using Fig. 6(b) was measured to be 11 ± 2 nm, which is close to the value obtained from HAADF-STEM image (in Fig. 5(b)). The lower value obtained could be due to misindexing at grain boundaries leading to blunt edges (refer to Fig. 6(b)).

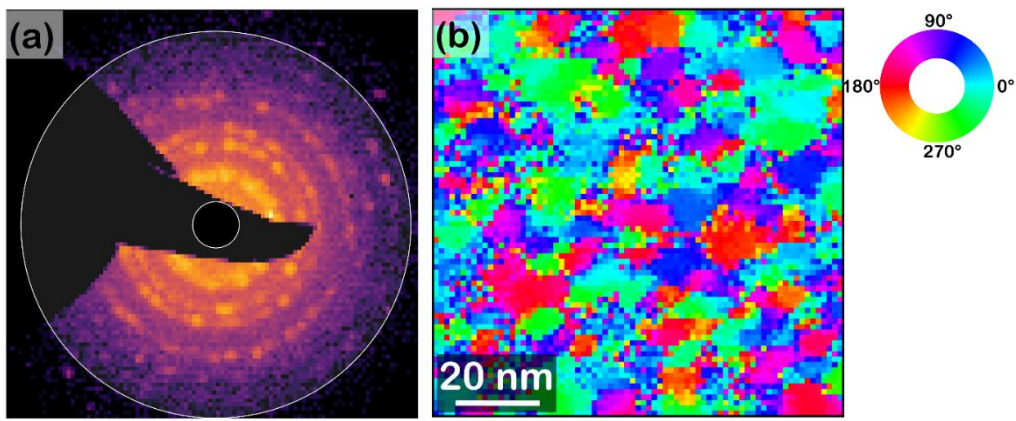

Fig. 6: (a) Average diffraction patterns from a 75 x 75 pixel cropped area of the sample scanned at a dwell time of 0.21 ms, and (b) In-plane orientation map.

## 4.2. Identifying twins in copper

The integration of 4D-STEM in SEM facilitates the capture of diffraction patterns, demonstrating possibilities as an alternative tool for TEM in various applications such as orientation mapping, strain analysis, or dislocation imaging. This integration not only enhances the throughput and versatility of SEM but also enables surface (topography) and internal (crystal structure) characterization within a single instrument, thereby reducing costs and promoting sustainability. To demonstrate this capability, a polycrystalline Cu sample containing annealing twins is fabricated using $Ga^+$ FIB and the SEM image is shown in Fig. 7(a). The zero-tilt configuration is employed with a working distance of 5 mm and a camera length of 19 mm. The data is collected at an acceleration voltage of 25 kV, a beam current of 500 pA and at a dwell time of 0.84 ms. The average of diffraction patterns collected is shown in Fig. 7(b). It can be inferred from the diffraction pattern that the matrix is away from any lower-order zone axis. The diffracted spots in the electron diffraction pattern were indexed, as presented in Fig. 7(b). Virtual dark-field (vDF) images were generated from the diffracted spots corresponding to (022) and (333), respectively, shown in Fig. 7(c) and (d).

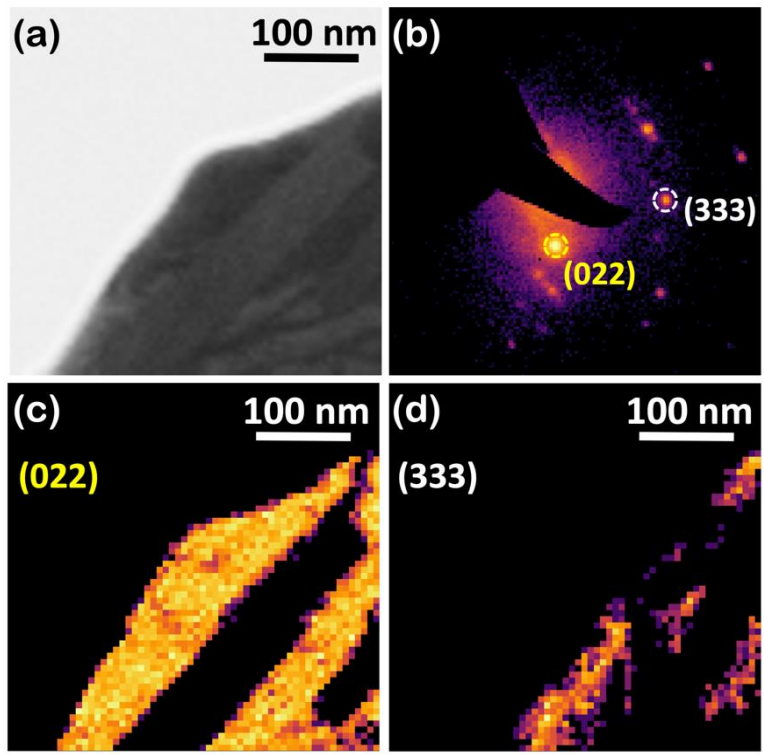

Fig. 7: (a) SEM image using ET detector of polycrystalline Cu, (b) average diffraction patterns obtained across the SEM image shown in (a), (c) virtual dark-field (vDF) image from the diffracted spot (022) showing the Cu grain, (d) vDF image from the diffracted spot (333) showing twins in the grain.

Radial Fourier analysis was carried out to determine the orientation relationship between two regions, as described in Section 4.1. From the region spanning over 30 to 50 mrads, the cartesian coordinates were converted to radial coordinates and in-plane orientation maps were constructed, as displayed in Fig. 8 (a). Across the interface of both regions, marked as a black arrow, the misorientation (in degrees) was measured and plotted with respect to distance. From the plot, a misorientation of approximately 62° was estimated, suggesting a twin boundary.

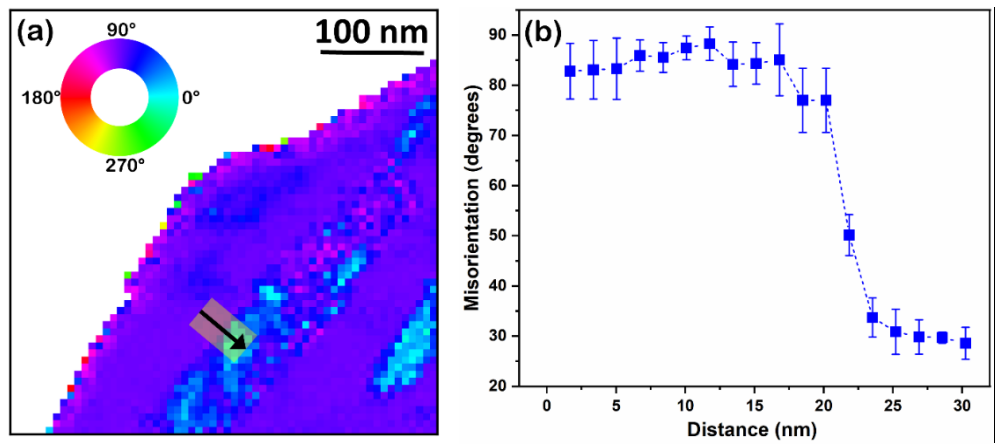

Fig. 8: (a) In-plane orientation map, and (b) Misorientation (in degrees) plot measured across the line in (b).

Further, it is to be noted that the twinning region in the vDF image in Fig. 7(d) lacks the desired contrast. Although contrast enhancement is feasible by tilting the sample to two-beam conditions, however, it is limited with our current setup. To further enhance the contrast, virtual dark-field images were generated using multiple diffracted spots corresponding to the Cu matrix and twinning region. Figure A3, in the appendix, displays the virtual dark-field image corresponding to each spot. In Fig. 9 (a), the diffracted spot corresponding to the Cu matrix is marked in yellow, and white for the twin region. The virtual dark-field images depicted in Fig. 9 (b) and (c) exhibit enhanced contrast compared to virtual dark-field images generated from single diffracted spots in Fig. 7. The addition of 4D-STEM in SEM enables the extraction of essential information about crystal structure and orientation from diffraction patterns, while also allowing the extraction of topographical details from the SE image, making it a powerful tool for comprehensive materials characterization. Furthermore, the crystallographic orientation analysis, such as in-plane orientation mapping shown in Fig. 6 or misorientation calculation in Fig. 8, can be further improved through diffraction pattern template matching, similar to automated crystal orientation mapping (ACOM) on TEM. We are currently developing a template-matching algorithm for our 4D-STEM data to enable 3D orientation mapping. However, this work is still in progress and is beyond the scope of this paper.

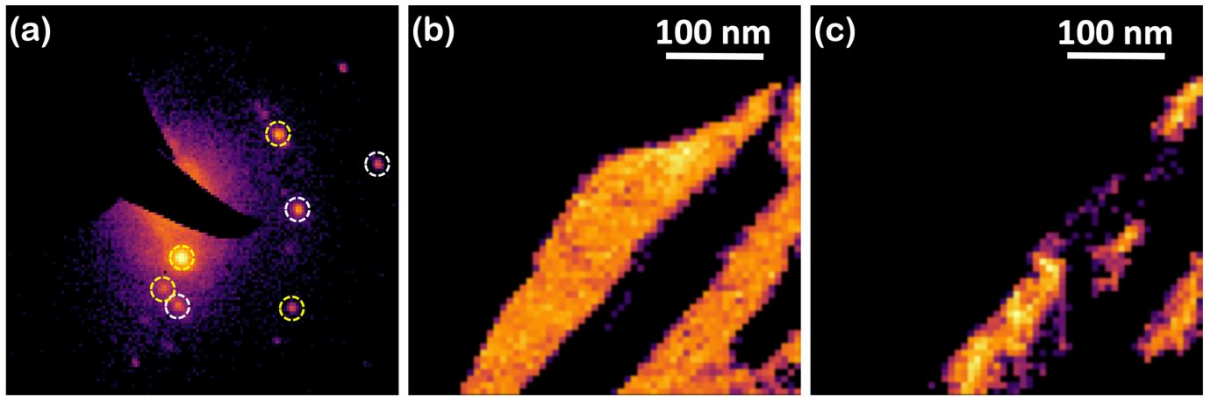

Fig. 9: (a) Average diffraction patterns overlapped with circles corresponding to Cu matrix (yellow circle) and twin (white circle), (b) and (c) virtual dark-field (DF) image from all the diffracted spots corresponding to Cu matrix and twins, respectively.

## 5. Conclusion

In conclusion, our study demonstrated the advancements in the implementation of 4D-STEM within SEM, showcasing its extended capabilities in several critical aspects. By utilizing the MiniPIX detector in event-driven mode, we achieved a notable increase in acquisition rate and a substantial reduction in data size, making 4D-STEM more feasible for *in situ* SEM testing. The stage-detector geometry allowed for an impressive camera length of 160 mm, enhancing angular resolution, and broadening its application showing its potentials in magnetic or electric field imaging. Additionally, our successful imaging of nanostructured Pt-Cu thin films and FIB-prepared polycrystalline Cu highlights the versatility of 4D-STEM or various materials, emphasizing its potential for detailed microstructural analysis. The level of resolution achieved using 4D-STEM in SEM not only underscores the synergistic benefits of combining in situ experiments with this technique but also paves the way for its broader application in materials science, providing a robust framework for future research and technological development.

**Appendix**

*A1. Indexing of diffracted rings in the diffraction pattern from Pt-Cu thin film sample*

**First approach – Direct measurements of visible rings:**

Figure A1(a) illustrates the diffraction data acquired from a Pt-Cu thin film in the zero-tilt configuration at a camera length of 19 mm. The measured distances, in pixels, of distinctly visible rings, marked in Figure A1(b), are provided in Table T1. The ratios between the distances of the second ring with the first, the third with the first, and so forth, were calculated. These ratios are crucial in identifying the planes, which are then indexed accordingly and documented in Table T1. The interplanar spacing were calculated using the lattice parameter of Pt ($a_{Pt}$ = 0.392 nm [27]). From the ratios and selection rule for FCC crystal structures [28], the indexed planes are (022), (113), (133), and (135). Furthermore, the value of one pixel based on interplanar spacing was determined to be ~0.352 nm$^{-1}$.

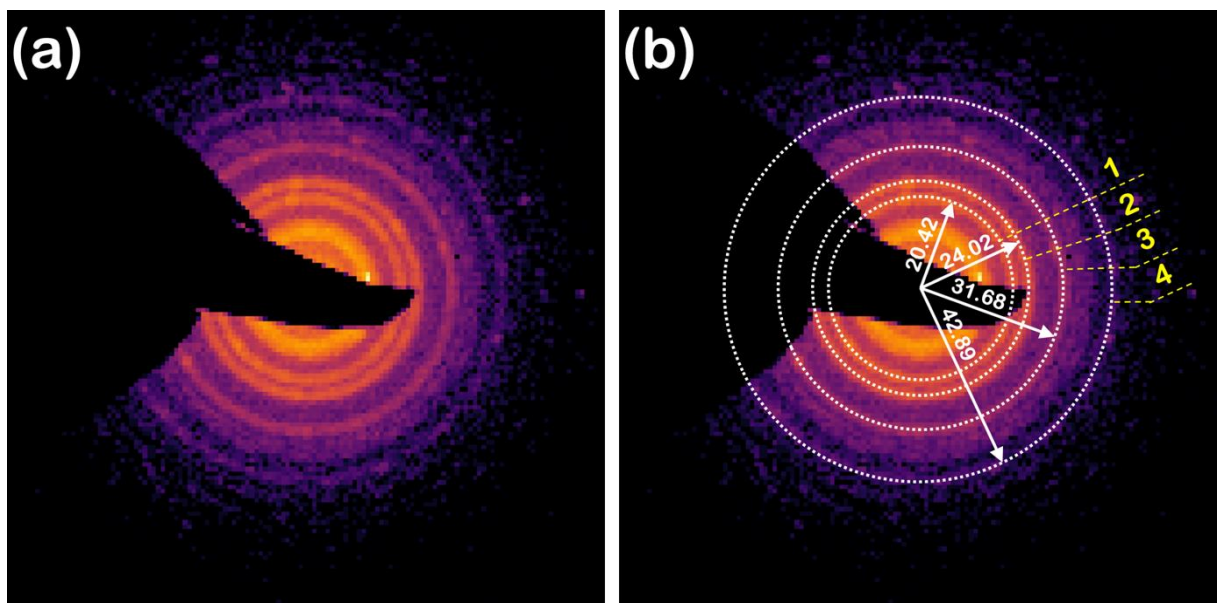

Fig. A1: Average diffraction patterns for Pt thin film overlapped with ring and length measured from origin in pixels.

**Table T1:** Lengths and ratios of distinctly visible rings in Pt-Cu thin film diffraction data, along with corresponding plane and interplanar spacing of Pt lattice.

| Ring Number | Length (in pixels) | Ratio with 1st ring | Possible plane (hkl) | Interplanar spacing (1/nm) | Ratio with (022) | Value of 1 pixel (1/nm) |
|---|---|---|---|---|---|---|
| 1 | 20.42 | 1.000 | (022) | 7.215 | 1.000 | 0.353 |
| 2 | 24.02 | 1.176 | (113) | 8.461 | 1.173 | 0.352 |
| 3 | 31.68 | 1.551 | (133) | 11.120 | 1.541 | 0.351 |
| 4 | 42.89 | 2.100 | (135) | 15.092 | 2.092 | 0.352 |

**Second approach – Fourier analysis across the ring-shaped region:**

In the approach mentioned above, certain peaks may remain undetermined due to poor signal-to-noise ratio. Hence, an alternate approach is adopted. Fourier analysis spanning from the center across the entire scan was conducted, and the resulting spectrum is depicted in Fig. A2(a). Subsequently, intensity curves were extracted, displayed in Fig. A2(b), and peaks were identified and indexed using the Pt database. This method, compared to the initial approach, allows for the determination of peaks corresponding to high-order planes.

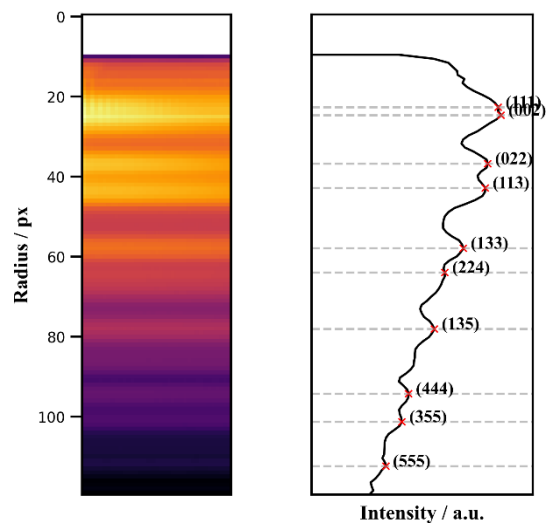

Fig. A2 (a) Spectrum spanning over an entire range, and (b) Intensity curve with peaks position marked and indexed using pure Pt-data base

*A2: Virtual dark-field (DF) images from individual diffracted spots in Cu sample*

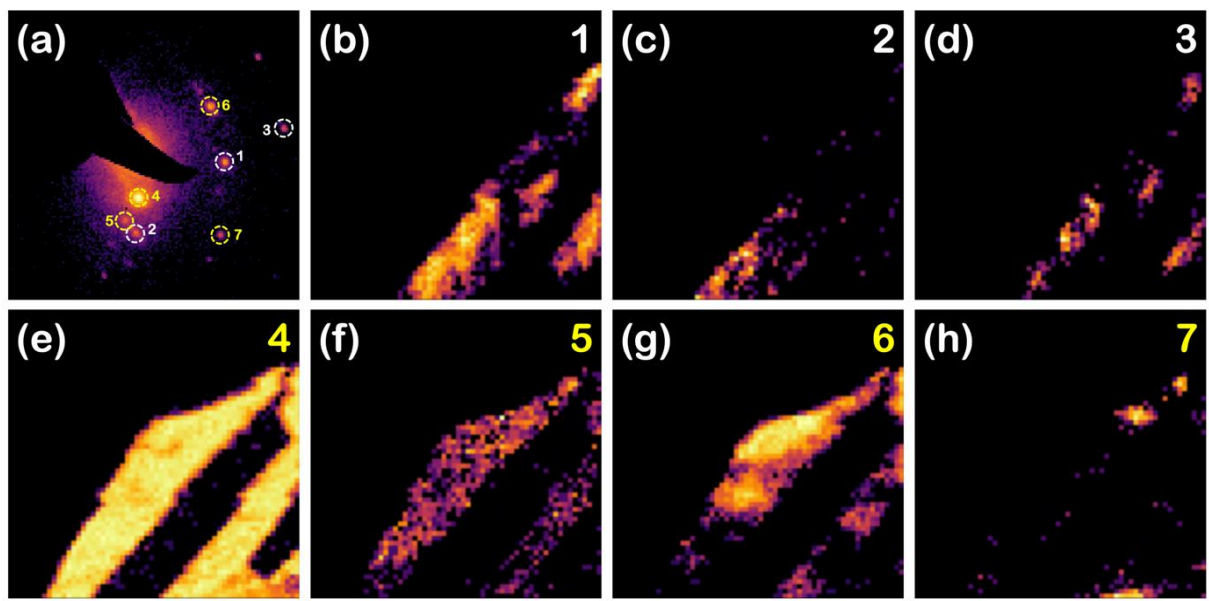

Fig. A3: (a) Average diffraction patterns, and (b-h) virtual dark-field (vDF) images from the diffracted spots marked in (a).

**CRediT authorship contribution statement**

**Ujjval Bansal:** Conceptualization, Methodology, Validation, Formal analysis, Investigation, Visualization, Writing – Original draft. **Amit Sharma:** Resources (Sample preparation) and Writing - Review & Editing. **Barbara Putz:** Resources (Sample preparation) and Writing - Review & Editing. **Christoph Kirchlechner:** Conceptualization, Funding acquisition, and Writing - Review & Editing. **Subin Lee:** Visualization, Supervision, Funding acquisition, and Writing - Review & Editing

**Declaration of competing interest**

The authors declare that they have no known competing financial interests or personal relationships that could have appeared to influence the work reported in this paper.

**Data availability**

Data available within the article.

**Acknowledgments**

We express our gratitude for the financial support provided by the Robert-Bosch-Foundation and the Helmholtz Program Materials Systems Engineering. Funding by the Helmholtz imaging platform of the project "Deep-learning assisted fast in situ 4D electron microscope imaging" is greatly acknowledged. We also acknowledge the support of the Karlsruhe Nano Micro Facility (KNMFi, www.knmf.kit.edu), a Helmholtz Research Infastriucture at Karlsruhe Institute of Technology, for facilitating the TEM work.